\documentclass[12pt]{article}

\usepackage[round,comma,authoryear]{natbib}
\usepackage{float}
\usepackage{amsmath, amssymb, amsfonts}
\RequirePackage{amsthm}
\usepackage{algorithm}
\usepackage{algorithmic}
\usepackage{hyperref}
\usepackage{amsmath, amssymb, amsfonts} 
\usepackage{graphicx, subfigure} 
\usepackage{booktabs}            
\usepackage{multirow}            
\usepackage{color,soul}
\usepackage{mdframed}
\usepackage{setspace}
\usepackage[utf8]{inputenc}
\usepackage{caption}
\usepackage{epstopdf}
\def\bSig\mathbf{\Sigma}

\newcommand{\bigCI}{\mathrel{\text{\scalebox{1.07}{$\perp\mkern-10mu\perp$}}}}
\graphicspath{{./figs/}}

\renewcommand\appendix{\par
  \setcounter{section}{0}
  \setcounter{subsection}{0}
  \setcounter{figure}{0}
  \setcounter{table}{0}
  \renewcommand\thesection{Appendix \Alph{section}}
  \renewcommand\thefigure{\Alph{section}\arabic{figure}}
  \renewcommand\thetable{\Alph{section}\arabic{table}}
}

\title{De novo construction of polyploid linkage maps using discrete graphical models}

\author{P. Behrouzi\\
  Wageningen University and Research Centre\\
  \texttt{pariya.behrouzi@wur.nl}
  \and E. C. Wit\\
  University of Groningen\\
  \texttt{e.c.wit@rug.nl}
  }

\date{}
\begin{document}

\maketitle

\begin{abstract}
Linkage maps are used to identify the location of genes responsible for traits and diseases. New sequencing techniques have created opportunities to substantially increase the density of genetic markers. Such revolutionary advances in technology have given rise to new challenges, such as creating high-density linkage maps. Current multiple testing approaches based on pairwise recombination fractions are underpowered in the high-dimensional setting and do not extend easily to polyploid species. To remedy these issues, we propose to construct linkage maps using graphical models either via a sparse Gaussian copula or a nonparanormal skeptic approach. 

We determine linkage groups (LGs), typically chromosomes, and the order of markers in each LG by inferring the conditional independence relationships among large numbers of markers in the genome. Through simulations, we illustrate the utility of our map construction method and compare its performance with other available methods, both when the data are clean and contain no missing observations and when data contain genotyping errors. 
Our comprehensive map construction method makes full use of the dosage SNP data to reconstruct linkage map for any bi-parental diploid and polyploid species.
We apply the proposed method to two genotype datasets: barley and potato from diploid and polyploid populations, respectively. 
The method is implemented in the R package {\tt netgwas} which is freely available at \url{https://cran.r-project.org/web/packages/netgwas}.

 \end{abstract}

\section{Introduction}
\label{chap3:into}

A linkage map provides a fundamental resource to understand the order of markers for the vast majority of species whose genomes are yet to be sequenced. Furthermore, it is an essential ingredient in the often used quantitative trait loci (QTL) mapping of genetic diseases, and particularly in identifying genes responsible for heritable or other types of diseases in humans or traits such as disease resistance in plants or animals.

Recent advances in sequencing technology make it possible to comprehensively sequence huge numbers of markers, construct dense maps, and ultimately create a foundation for studying genome structure and genome evolution, identifying QTLs and understanding the inheritance of multi-factorial traits. Next--generation sequencing (NGS) techniques offer massive and cost--effective sequencing throughput. However, they also bring new challenges for constructing high--quality linkage maps. NGS data can suffer from high rates of genotyping errors, as the observed genotype for an individual is not necessarily identical to its true genotype. Under such circumstances, constructing high--quality linkage maps can be difficult. 

Each species is categorized as diploid or polyploid by comparing its chromosome number. Diploids have two copies of each chromosome. For diploid species many algorithms for constructing linkage maps have been proposed. 
Some of them have been implemented into user-friendly software, such as R/qtl \citep{broman2003r}, J{\scriptsize OIN}M{\scriptsize AP} \citep{jansen2001constructing}, OneMap \citep{margarido2007onemap}, and MST{\scriptsize MAP} \citep{wu2008efficient}. 
Among the algorithms for constructing genetic maps, R/qtl estimates genetic maps and identifies genotyping errors in relatively small sets of markers. J{\scriptsize OIN}M{\scriptsize AP} is a commercial software widely used in the scientific genetics community. It uses two methods to construct genetic maps: one is based on regression \citep{stam1993construction} and the other uses a Monte Carlo multipoint maximum likelihood \citep{jansen2001constructing}. OneMap has been reported to construct linkage maps in non-inbred populations. However, it is computationally expensive. The MSTMap is a fast genetic map algorithm that determines the order of markers by computing the minimum spanning tree of an associated graph. 

Polyploid organisms have more than two chromosome sets. Polyploidy is very common in flowering plants and in different crops such as watermelon, potato, and bread wheat, which contain three (triploid), four (tetraploid), and six (hexaploid) sets of chromosomes, respectively. Despite the importance of polyploid species, statistical tools for construction of their linkage map are underdeveloped. However, \cite{grandke2017pergola} recently developed a method for this purpose. Their method is based on calculating recombination frequencies between marker pairs, then using hierarchical clustering and an optimal leaf algorithm to detect chromosomes and order markers. Nevertheless, this method can be computationally expensive even for a small numbers of markers. PolymapR \citep{bourke2017polymapr} is another software that construct a genetic map from bi-parental populations of outcrossing autopolyploids. It clusters markers over a range of LOD thresholds; it requires users to select the LOD threshold that best clusters the data. Then it uses weighted linear regression or multi-dimensional scaling for ordering markers. Most literature has focused on constructing genetic linkage maps for tetraploids, but these are limited only to autotetraploid species. TetraploidSNPMap, is a software for this situation \citep{hackett2017tetraploidsnpmap}, but because it needs manual interaction and visual inspection its application is limited. For example, user needs to specify how many linkage groups (chromosomes) the algorithm should be detected. 
Furthermore, current approaches to polyploid map construction are based mainly on estimation of recombination frequency and LOD scores \citep{wang2016potential}, which does not use the full multivariate information in the data.

Different diploid and polyploid map construction methods have made substantial steps toward building better--quality linkage maps. However, the existing methods still suffer from low quality genetic mapping performance, in particular when ratios of genotyping errors and missing observations are high. The main contribution of this paper is to introduce, for both diploid and polyploid species, a novel linkage map algorithm to overcome the difficulties arising routinely in NGS data. With the proposed method we aim to build high--density and high--quality linkage maps using the statistical property called conditional dependence relationships, which reveals direct relations among genetic markers. For diploid scenarios, we evaluated the performance of the proposed method and the other methods in several comprehensive simulation studies, both when the input data were clean and had no missing observations and when the input data were very noisy. 
We measured the performance of the methods in accuracy scores of grouping and ordering. In addition, we studied the performance of our method in constructing linkage maps for simulated polyploids, namely tetraploids and hexaploids. Furthermore, we applied the map construction method in {\tt netgwas} \citep{behrouzi2017netgwas} to construct maps for two genotype datasets: barley and potato from diploid and tetraploid populations, respectively.    
\begin{figure}[t]
\centering
\includegraphics[width=0.98\textwidth]{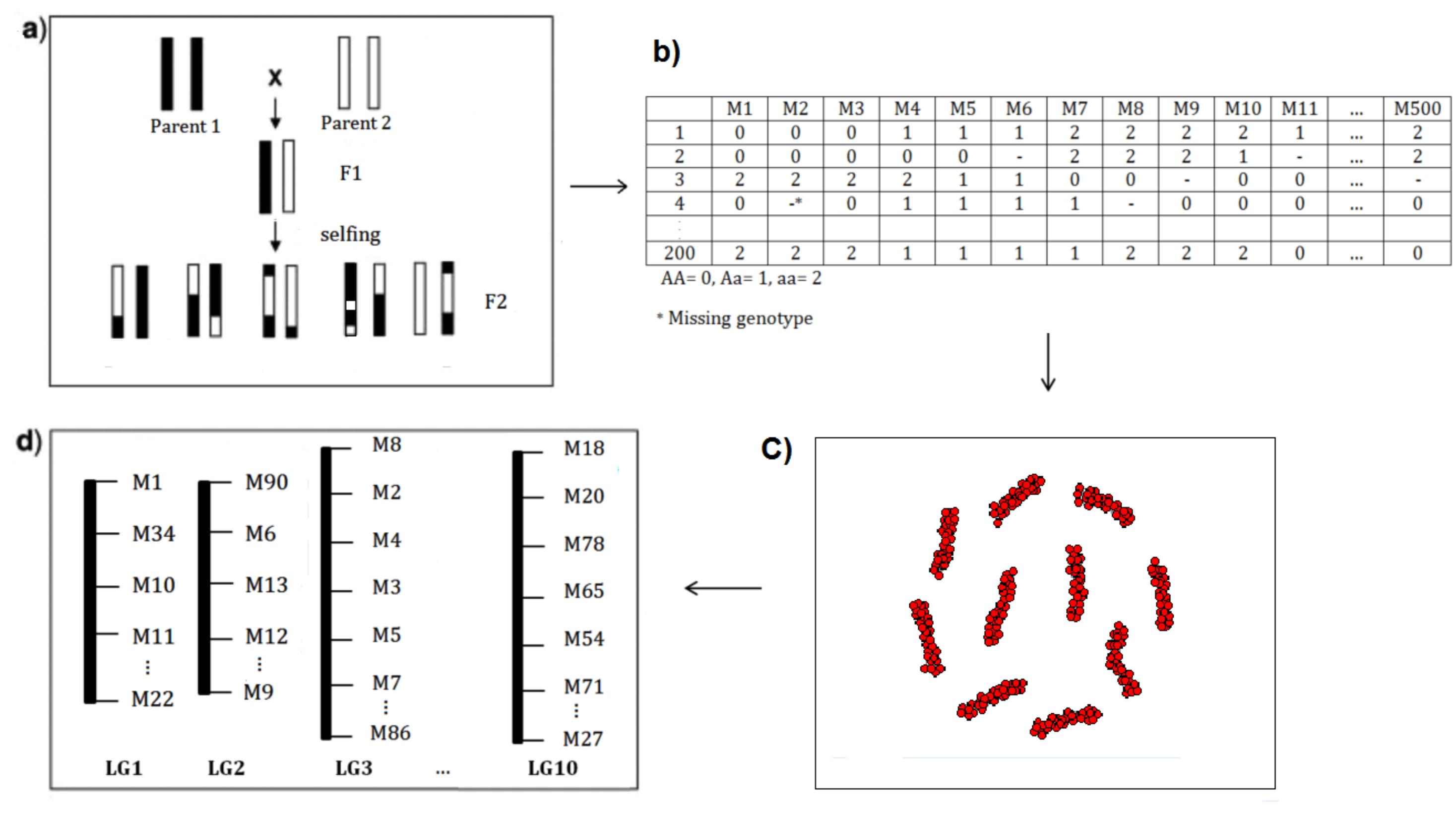}
\caption{General view of proposed linkage map estimation process. To illustrate, we use a diploid population containing two copies of each chromosome. (a) Example of mating experiment of an inbred F2 population. (b) Derived genotype data for $200$ individuals which have genotyped for $500$ markers. (c) Reconstruction of undirected graph between all $500$ markers. (d) $10$ linkage groups (chromosomes) with markers ordered within each linkage group (LG).}
\label{process}
\end{figure}
\section{Genetic background on linkage map}
A linkage map is the linear order of genetic markers on a chromosome. Geneticists use it to study the association between genes and traits. In this section we describe the relationship between a linkage map and single nucleotide polymorphism (SNP) markers. For the moment, we assume that each allele can take only one of two values, $A$ or $a$. This assumption can be relaxed without requiring any methodological adjustments; more will follow in the discussion. Here, we are dealing with markers from high--throughput data such as NGS and SNP arrays. 
\subsection{Linkage map for diploids and polyploids}
\label{qploid}
Diploid organisms contain two sets of chromosomes, one from each parent, whereas polyploids contain more than two sets of chromosomes. In polyploids the number of chromosome sets reflects their level of ploidy: triploids have three sets, tetraploids have four, pentaploids have five, and so forth. 
Here, we refer to diploids and polyploids as q-ploid $q \ge 2$, where in diploids $q = 2$, triploids $q = 3$, tetraploids $q=4$, and so on.

The genotype of any q-ploid organism can be homozygous or heterozygous at each single locus on the genome. Different genotype forms of the same gene are called alleles. Alleles can lead to different traits. Alleles are commonly represented by letters; for example, for the gene related to the trait, the allele could be called A and a. In q-ploid individuals there are q copies of allele. If all q allele copies of an organism are identical, the organism is in the homozygous state at that locus; otherwise it is in the heterozygous state. For instance, a tetraploid individual is homozygous for two size alleles, A and a, if all 4 allele copies are either $A$, or $a$, which correspond with the genotypes $AAAA$ and $aaaa$, respectively. If a tetraploid individual is heterozygous the following three genotypes would appear: one copy of the A allele and three copies of a (e.g. Aaaa), two copies of A and two copies of a (e.g. AAaa), or three copies of A and one copy of a (e.g. AAAa). Unlike existing methods, our method works not only for diploid organisms but also for all polyploids. Obviously, our method can also be used to analyze simple haploid organisms such as haploid yeast cells. 
\subsection{Mapping population} 
\label{inbredpop}
Mating between two parental lines with recent common biological ancestors is called inbreeding. Mating between parental lines with no common ancestors up to e.g. $4$-$6$ generations is called outcrossing. In both cases, the genomes of the derived progenies are random mosaics of the genomes of the parents. As a consequence of inbreeding parental alleles are attributable to each parental line in the genome of the progeny, whereas in outcrossing this is not the case. 

Inbreeding progenies derive from two homozygous parents. Some inbreeding designs, such as \emph{Backcrossing} (BC), lead to a homozygous population where the derived genotype data include only homozygous genotypes of the parents, namely AA and aa (conveniently coded as $0$ and $1$). However, some other inbreeding designs such as $F2$ lead to a heterozygous population, where the derived genotype data contain both heterozygous and homozygous genotypes, namely AA, Aa, and aa (conveniently coded as $0$, $1$ and $2$; see Figures \ref{process}a and \ref{process}b for an example of a diploid species). Although many other experimental designs are being used in genetic studies, not all existing methods for linkage mapping support all inbreeding experimental designs. However, our proposed algorithm constructs a linkage map for any type of biparental inbreeding experimental designs. In fact, unlike other existing methods, our approach does not require specifying the population type because it is broad and handles any population type that contains at least two distinct genotype states.

Outcrossing or outbred experimental designs, such as full--sib families, derive from two non--homozygous parents. Thus the genome of the progenies includes a mixed set of many different marker types, including fully informative markers and partially informative markers (e.g. missing markers). Markers are called fully informative when all of the resulting gamete types can be phenotypically distinguished on the basis of their genotypes; they are called partially informative when the gamete types have identical phenotypes. 
\begin{figure}[t]
\centering
\includegraphics[width=0.98\textwidth]{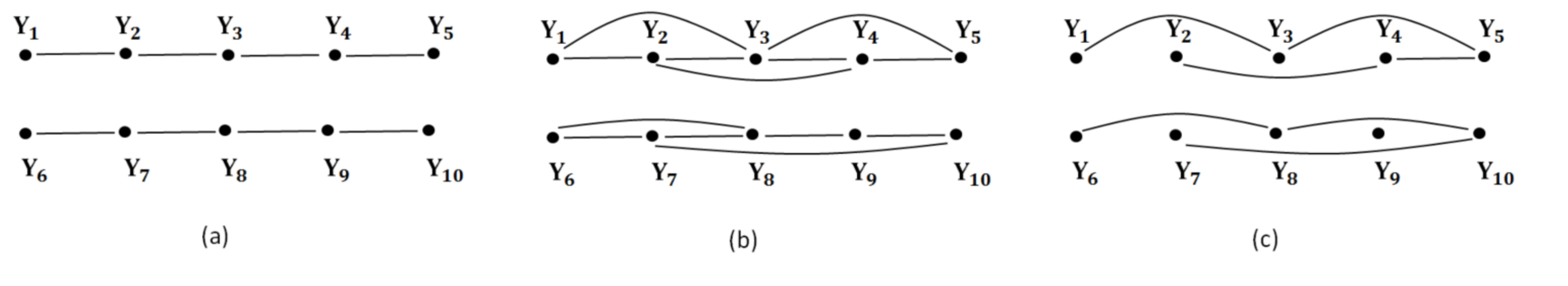}
\caption{Cartoon example of conditional dependence pattern between neighboring markers in different population types: (a) homozygous, (b) inbred, (c) outcrossing (outbred) populations, where ordered markers $Y_1, \ldots, Y_5$ reside on chromosome $1$, and $Y_6, \ldots, Y_{10}$ on chromosome $2$. }
\label{schemeCI}
\end{figure}
\subsection{Meiosis and Markov dependence}
\label{meioisMD}
During meiosis, chromosomes pair and exchange genetic material (crossover). In diploids, pairing at meiosis occurs between two chromosomes. In polyploids the $q$ chromosome copies may form different types of multivalent pairing. For example, in tetraploids all four chromosome copies may pair at meiosis. 
Assume a sequence of ordered SNP markers $X^c_1, X^c_2, \ldots, X^c_d$ along chromosome $c$ in a q-ploid species. We describe the Markov dependence structure between markers for different population schemes. (i) During meiosis in inbred populations, genetic material from one of the two parents is copied into the offspring in a sequential fashion, i.e. reading along the genome, until the copying switches in a random fashion to the other parent. Thus, the genome of the offspring is a random but piecewise continuous mosaic of the genomes of its parents. The genotype state at each chromosomal region, or locus, of the offspring is either homozygous maternal, heterozygous, or homozygous paternal. For instance, as a result of genetic linkage and crossover a homozygous maternal genotype will typically be followed by a heterozygous genotype before being able to be followed by a homozygous paternal genotype. 

\emph{Genetic linkage} means that markers located close to one another on a chromosome are linked and tend to be inherited together during meiosis. Another key biological fact is that during meiosis markers on different chromosomes segregate independently; this is called the \emph{independent assortment law}.  

For example, in scheme (i) consisting of only a homozygous population, the random variable $Y_j$ which represents the genotype of an individual at location $j$ can be defined as 
\begin{eqnarray}
Y_{j} = \left\{
\begin{array}{l l}
	1  & \mbox{paternal marker at locus $j$ on homologue k,} \\
	0  & \mbox{otherwise.} \nonumber
\end{array} \right.
\end{eqnarray}

This scheme occurs in inbred homozygous populations that include only two genotype states, namely homozygous maternal and homozygous paternal. Mapping populations, such as backcrossing, are included in this scheme. Then, under the assumption of no crossover interference -- meaning when a crossover has formed, other crossovers are not prevented from forming -- the recombination frequency between the two locations $j$ and $j+1$ is independent of recombination at the other locations on the genome. So, the following holds 
\begin{equation}
\label{markovInbred}
Pr(Y_{j+1}= y_{j+1} \ | \ Y_{j}= y_{j} , Y_{j-1}= y_{j-1}, \ldots, Y_{1}= y_{1} ) = Pr(Y_{j+1}= y_{j+1} \ | \ Y_{j}= y_{j})
\end{equation}
This equation indicates that the genotype of a marker at location $j+1$ is conditionally independent of genotypes at locations $j-1, j-2, \ldots, 1$ given a genotype at location $j$. This can be written as 
\begin{equation}
\centering
\label{CI}
Y_{j + 1} \bigCI  (Y_1, \ldots, Y_{j-1}) \ | \ Y_j
\end{equation}
This defines a discrete graphical model $G=(V, E)$ which consists of vertices $V= \{1, \ldots, p\}$ and edge set $E \subseteq V \times V$ with a binary random variable $Y_j \in \{0,1\}^p$. Given the above property between neighboring markers, we construct linkage maps using conditional (in)de\-pen\-den\-ce models. Figure \ref{schemeCI}a shows a cartoon image of conditional (in)de\-pen\-den\-cies for this scheme.

Scheme (ii): In inbred populations, one complication arises when in the genotype data we cannot identify each homologue due to heterozygous genotypes. Q-ploid ($q \ge 2$) heterozygous inbred populations, like $F2$, are examples of such cases, where we define $X_{jk}$ as 
\begin{eqnarray}
X_{jk} = \left\{
\begin{array}{l l}
	1  & \mbox{if marker $j$ on homologue k is of type $A$,} \\
	0  & \mbox{otherwise} \nonumber
\end{array} \right.
\end{eqnarray}
where A is one of the two possible alleles at that specific location. Here, $X_{jk}$ represents the allele at homologue $k$ of a chromosome, where the genotype in that location can be written as $X_{j.} = \{X_{j1} \ldots X_{jq}\}$. For example, at marker location $j$, $X_j= Aaaa$ is one possible genotype for a tetraploid species ($q = 4$); it includes one copy of the desirable allele $A$ where $X_{j1}= 1$, $X_{j2}= 0$, $X_{j3}= 0$, and $X_{j4}= 0$ represent the alleles in the first, second, third and fourth homologues, respectively. The other possible genotypes which include one copy of the desired allele $A$ are $aAaa$, $aaAa$, $aaaA$. Because it is typically impossible to distinguish between genotypes with the same number of copies of a desired allele (e.g. $Aaaa$, $aAaa$, $aaAa$, $aaaA$), we therefore take a random variable $Y_j$ as observed in the number of $A$ alleles at location $j$: 
\begin{equation}
\label{XYFormula}
Y_j = \sum\limits_{k=1}^{q} X_{jk}.
\end{equation}
Table \ref{XY} shows an example of correspondence between $Y_j$ and $X_{j.}$ for a q-ploid species when $q=4$. We note that a q-ploid species contains $q+1$ genotype states at location $j$, as shown in Table \ref{XY} for a tetraploid species.

Due to \emph{genetic linkage}, the sequence of ordered SNP markers $Y_1, Y_2, \ldots, Y_{d}$ forms a Markov chain as equation (\ref{markovInbred}) with state space $S$ which contains $q+1$ states. Therefore, the conditional (in)dependence relationship (\ref{CI}) between neighboring markers is held. Figure \ref{schemeCI}b presents a cartoon image of the conditional independence graph for this scheme.
\begin{table}[h]
\centering
\caption{Number of copies (dosage) of a reference allele. Relation between different genotypes, $X_{j.}$, and allele dosage, $Y_j$, for a tetraploid individual, where $A$ is the reference allele.} 
\begin{tabular}{ |p{1.3cm}|p{8cm}| }
\hline
  $Y_j$		& $X_{j.}$ \\
 \hline
 \vspace{-0.1cm}
 $0$		&  \vspace{-0.1cm} $aaaa$ \\
 $1$		& $Aaaa$, $aAaa$, $aaAa$, $aaaA$ \\
 $2$		& $AAaa$, $AaAa$, $AaAa$, $AaaA$, $aaAA$ \\
 $3$		& $AAAa$, $AaAA$, $AAaA$, $AAAa$, $aAAA$ \\
 $4$		& $AAAA$					\\
 \hline
\end{tabular}
\label{XY}
\end{table}
Scheme (iii): In outcrossing (outbred) populations, unlike inbred populations, the meaning of ``parental" is either unknown or not well defined. In other words, markers in the genome of the progenies can not  easily be assigned to their parental homologues. For example, if both non-homozygous parents contain $A_jA_jA_jA_j$ genotype at marker location $j$, then offspring will also have $A_jA_jA_jA_j$ genotype at marker location $j$. But we do not know whether that genotype belongs to the paternal or maternal homologue, since both parents have $A_jA_jA_jA_j$ genotype at marker location $j$. So, in this case we define  $X_{jk}$ as follows
\begin{eqnarray}
X_{jk} = \left\{
\begin{array}{l l}
	1  & \mbox{if marker $j$ on homologue $k$ is of type $A_j$,} \\
	0  & \mbox{otherwise} \nonumber
\end{array} \right.
\end{eqnarray}
where $A_j$ is one of the possible parental alleles at location $j$. So, random variable $Y_j$ which represents the dosage of alleles, can be defined as equation (\ref{XYFormula}). 

Furthermore, in polyploids the \emph{linkage} depends on how a single chromosome pairs during meiosis to generate gametes. In this regard, if both polyploid parents have an $A_j$ allele in all $q$ haploids, then the offspring will also have it, and this will not co-vary with neighboring markers. The possibility of different pairing models during meiosis makes the situation more complex. In diploids, the two homologue chromosomes pair up and form a bivalent, then cross-over before recombinations occur. But polyploid meiosis can occur in various ways; in tetraploids four homologue chromosomes can during meiosis form either two separate bivalents, each of which contributes one haploid, like diploids, or, alternatively, in a more complex situation, the four homologue chromosomes can form quadrivalents, so that cross-over occurs between eight haploids. In both pairing models, bivalent or quadrivalent, crossover events result in recombined haploids that are mosaics of parental chromosomes. Outbred progenies are genetically diverse and highly heterozygous, whereas inbred individuals have little or no genetic variation. 

The term (\ref{markovInbred}) partially holds for the scheme (iii), where a discrete graphical model can be defined for a multinomial variable $Y_j = \{ 0, 1, \ldots, q \}$. We use conditional independence to construct linkage maps in outbred populations. However, in this type of population, due to a mixed set of different marker types, the conditional independence relationship between neighboring markers may be more complicated. Many genetic assumptions made in traditional linkage analyses (e.g., known parental linkage phases throughout the genome) do not hold here. For example, when both parents have $A_j$ allele, then their offspring will also have it; however this will not covary with neighboring markers. Figure \ref{schemeCI}c shows a cartoon example of such conditional independence graphs.

To summarize, term (\ref{markovInbred}) holds for schemes (i) and (ii), and partially (iii) because transition probability from a genotype at location $j$ to a genotype at location $j+1$ depends on the recombination frequency between the two locations $j$ and $j+1$, which is independent of recombination in the other locations. This can be modeled by a discrete Markov process $\{Y_j\}_{j=1,\ldots,d}$ with state space $S$ which contains  $q + 1$ genotype states and a transition matrix, which, in case of polyploids ($q \ge 3$), can be calculated with respect to the mode of chromosomal pairing (e.g. bivalent or quadrivalent). 
The Markov structure of the SNP markers in all three schemes yields a graphical model with as many nodes as markers in a genome. The random variable $Y_j$ follows a discrete graphical model whereby the joint distribution $P(Y)$ can be factorized as,
\begin{equation}
\label{eq1}
P(Y)= \prod_{c=1}^{C} \prod_{j=1}^{p_c-1} f^{(c)}_{j, {j + 1}}(Y^{(c)}_{j} ,Y^{(c)}_{j + 1}), 
\end{equation}
where $C$ defines the number of chromosomes in a genome, and $p_c$ stands for the number of markers in chromosome $c$. The outer multiplication of (\ref{eq1}) shows the \emph{independent assortment law}, and the inner multiplication represents the \emph{genetic linkage} between markers within a chromosome, where the factor $f^{(c)}_{j,j+1}$ indicates the conditional dependence between adjacent markers, given the rest of the markers. Through this probabilistic insight, the inferred conditional (in)dependence relationship between markers provides a high-dimensional space for the construction of a linkage map.
\section{Algorithm to detect linkage map}
\label{chapter4:method}
We propose to build a linkage map in two steps; first, we reconstruct an undirected graph for all SNP markers on a genome, and second, we determine the correct order of markers in the obtained linkage groups from the first step. We also show how our method handles genotyping errors and missing observations in reconstructing a linkage map. 
\subsection{Estimating marker-marker network}
\label{3Dmap}
To reconstruct an undirected graph between SNP markers in a q-ploid species we propose two methods: the sparse ordinal glasso approach \citep{behrouzi2017detecting} and the nonparanormal skeptic approach \citep{liu2012high} (the latter discussed under Supplementary Materials). The former method can deal with missing values, whereas the latter is computationally faster.

An undirected graphical model for the joint distribution (\ref{eq1}) of a random vector $Y = (Y_1, \ldots, Y_p)$ is associated with a graph $G = (V, E)$, where each vertex $j$ corresponds to a variable $Y_j$. The pair $(j, l)$ is an element of the edge set $E$ if and only if $Y_j$ is dependent of $Y_l$, given the rest of the variables. In the graph estimation problem, we have $n$ samples of the random vector $Y$, and it is our aim to estimate the edge set $E$. Depending on how various mapping populations are produced, $Y$ represents either binary variables $Y = \{0,1\}$, as in homozygous populations, or multinomial variables $Y= \{ 0, 1, \ldots, q + 1\}$ where $q$ is the ploidy level. For example in diploids $q$ is $2$ and in tetraploids $4$.

\paragraph{Sparse ordinal glasso} A relatively straightforward approach to discover the conditional (in)dependence relation among markers is to assume underlying continuous variables $Z_1, \ldots, Z_p$ for markers $ Y_1, \ldots, Y_p$, which can not be observed directly. In our modeling framework, $Y_j$ and $Z_j$ define observed rank and true latent value, respectively, where each latent variable corresponds to one observed variable. The relationship between $Y_j$ and $Z_j$ is expressed by a set of cut-points $(-\infty, C^{(j)}_1], (C^{(j)}_1, C^{(j)}_2], \ldots, \allowbreak (C^{(j)}_q, \infty)$, which is obtained by partitioning the range of $Z_j$ into $q_j - 1$  disjoint intervals. Thus, $y_j^{(i)}$, which represents the genotype of the $i$-th sample for the $j$-th marker, can be written as follows 
\begin{eqnarray}
y_j^{(i)} = \sum_{k= 1}^{q} k \times 1_{ \{ C^{(j)}_{q-1} < z_j^{(i)} \le C^{(j)}_{q} \}} 	\qquad i= 1, 2,\ldots, n,		
\end{eqnarray}
where we define $\mathcal{D}= \{z_j^{(i)} \in \mathbb{R} \ | \  C^{(j)}_{q-1} < z_j^{(i)} \le C^{(j)}_{q} \} $. We use a high dimensional Gaussian copula with discrete marginals. 
We assume 
\[
Z \sim N_p (0, \Sigma)
\]
where the $p \times p$ precision matrix $\Theta = \Sigma^{-1}$ contains all the conditional independence relationships between the latent variables.  
Given our parameter of interest $\Theta$, we non-parametrically estimate the cut-points for each $j= 1, \ldots, p$ as follows
\begin{eqnarray}
\widehat{C}^{(j)}_q = \left\{
\begin{array}{l l}
	-\infty  & \mbox{if  $q = 0$ ;} \\
	\Phi^{-1}(\sum_{i=1}^{n} I(y_j^{(i)} \leq q) /n) & \mbox{if $q= 1$, \ldots, $q_{j}-1$;} \\
	+\infty  & \mbox{if $q= q_j$.}
\end{array} \right. \nonumber
\end{eqnarray}

\paragraph{Penalized EM algorithm} In genotype datasets we commonly encounter situations where the number of genetic markers $p$ exceeds the number of samples $n$. To solve this dimensionality problem we propose to impose an $l_1$ norm penalty on the likelihood consisting of the absolute value of the elements of the precision matrix $\Theta$. Furthermore, to be able to deal with commonly occurring missing values in genotype data we implement an EM algorithm \citep{mclachlan2007algorithm}, which iteratively finds the penalized maximum likelihood estimate $\widehat{\Theta}_\lambda$. This algorithm proceeds by iteratively computing the conditional expectation of complete log-likelihood and optimizing it. 
In the E-step we compute the conditional expectation in the penalized log-likelihood
\begin{align}	
\label{E-step}
Q_\lambda(\mathbf{ \mathbf{\Theta}} \  | \ \widehat{\mathbf{ \mathbf{\Theta}}}^{(m)}) = & \frac{n}{2} \bigg[\log  |   \mathbf{ \mathbf{\Theta}}  |  -tr(\frac{1}{n} \sum_{i=1}^{n} E_{Z^{(i)}}(Z^{(i)} Z^{(i)t}  |  y^{(i)}, \widehat{ \mathbf{ \mathbf{\Theta}}}^{(m)}, \widehat{\mathcal{D}} )  \mathbf{ \mathbf{\Theta}}) -p \log(2\pi)\bigg] \\ \nonumber
- & \lambda || \Theta ||_1  
\end{align}
where $\lambda$ is a nonnegative tuning parameter. To calculate the conditional expectation $\bar{R} = \frac{1}{n} \sum_{i=1}^{n} E_{Z^{(i)}}(Z^{(i)} Z^{(i)t}  |  y^{(i)}, \widehat{\Theta}^{(m)}, \widehat{\mathcal{D}})$ we propose two different approaches, namely Gibbs sampling and an approximation method \citep{behrouzi2017detecting, guo2015graphical}. Further details on the calculation of the conditional expectation are provided in the Supplementary Materials. The M-step is a maximization problem which can be solved efficiently using either graphical lasso \citep{friedman2008sparse}
\begin{equation}
\label{glasso}
\widehat{ \mathbf{\Theta}}^{(m+1)}_{glasso}= \arg \max_ \mathbf{\Theta} \bigg\{\log | \mathbf{\Theta}| - tr(\bar{R}  \mathbf{\Theta}) - \lambda ||  \mathbf{\Theta}||_1 \bigg\} \nonumber
\end{equation}
or the CLIME estimator \citep{cai2011constrained} 
\begin{equation}
\label{CLIME}
\widehat{\Theta}^{(m+1)}_{\mbox{\tiny CLIME}}= \arg \min_\Theta ||\Theta||_1 \qquad \mbox{subject to} \qquad ||\bar{R}\Theta - I_p||_{\infty} \leq \lambda, \nonumber
\end{equation} 
where $I_p$ is a p-dimensional identity matrix.

In large-scale genotyping studies, it is common to have missing genotype data. Before determining the number of linkage groups and ordering markers, we handle the missing data within the E-step of the EM algorithm, where we calculate the conditional expectation of true latent variables given the observed ranks. If an observed value, $y_j^{(i)}$ is missing, we take the unconditional expectation of the corresponding latent variable. In the EM framework we can easily handle high ratios of missingness in the data. 
\subsection{Determining linkage groups}
\label{cluster}
A group of loci that are correlated defines a linkage group (LG). Depending on the density and proximity of the underlying markers each LG corresponds to a chromosome or part of a chromosome.  The number of discovered linkage groups is controlled by the tuning parameter $\lambda$ (section \ref{3Dmap}). We use the extended Bayesian criterion (eBIC), which has successfully been applied by \cite{yin2011sparse} in selecting sparse Gaussian graphical models for genomic data to determine the number of linkage groups. The eBIC is defined as 
\begin{equation}
\label{ebic}
eBIC (\lambda)= - 2 \ell(\widehat{\Theta}_{\lambda}) + (\log n + 4 \gamma \log p ) \mbox{df}(\lambda),
\end{equation}
where $\ell(\widehat{\Theta}_\lambda)$ is the non-penalized likelihood and $\gamma \in [0, 1]$ is an additional parameter. And $df(\lambda) = \sum_{1 \le i < j \le p} I(\widehat{\theta}_{ij,\lambda} \ne 0)$ where $\widehat{\theta}_{ij,\lambda}$ is $(i,j)$th entry of the estimated precision matrix $\widehat{\Theta}_\lambda$ and $I$ is the indicator function. In case of $\gamma= 0$ the classical BIC is obtained. Typical values for $\gamma$ are $1/2$ and $1$. We select the value of $\lambda$ that minimizes (\ref{ebic}) for $\gamma =\frac{1}{2}$. We note that in practice there is an opportunity that linkage groups have been selected manually given a prior knowledge. 

It is notable that in existing map construction methods the construction of linkage groups is usually done by manually specifying a threshold for pairwise recombination frequencies; this, however, influences the output map, whereas our method detects linkage groups automatically in a data--driven way.

Figure \ref{process}(c) shows an example of an estimated conditional independence graph between markers. This graph includes $10$ distinct sub--graphs, each of which corresponds to a linkage group. In this graph, given all markers on a genome, markers within the linkage groups are conditionally dependent, due to \emph{genetic linkage}, and markers between linkage groups are conditionally independent, due to the \emph{independent assortment law}.

Some genotype studies suffer from low numbers of samples or they contain signatures of epistatic selection \citep{behrouzi2017detecting}, which may cause bias in determining the linkage groups. To address this problem, besides the model selection step, we use the fast-greedy algorithm to detect the linkage groups in the inferred graph. This community detection algorithm reflects the two biological concepts of genetic linkage and independent assortment in a sense that it defines communities which are highly connected within, and have few links between communities.
\subsection{Ordering markers}
\label{ordeing}
Assume that a set of $d$ markers has been assigned to the same linkage group. Let $G(V^{(d)},E^{(d)})$ be a sub--graph on the set of unordered $d$ markers, where $V^{(d)} = \{1, \ldots, d\}$, $d \leq p$ and the edge set $E^{(d)}$ represents the estimated edges among $d$ markers where $E^{(d)} \subseteq E$. We remark that the precision matrix $\widehat{\Theta}_\lambda^{(d)}$, a submatrix of $\widehat{\Theta}_\lambda$, contains all conditional dependence relations between the set of $d$ markers. Depending on the type of mating between the parental lines we introduce two methods to order markers, one based on dimensionality reduction and another based on bandwidth reduction. Both methods result in a one-dimensional map.
\paragraph{Inbred} In inbred populations, markers in the genome of the progenies can be assigned to their parental homologues, resulting in a simpler conditional independence pattern between neighboring markers. In the case of inbreeding, we use multidimensional scaling (MDS) to represent the original high-dimensional space in a one-dimensional map while attempting to maintain pairwise distances. We define the distance matrix $D$ which is a $d \times d$ symmetric matrix where $D_{ii} = 0$ and $D_{ij} = - \log (\rho_{ij})$ for $i\ne j$. Here, the matrix $\rho$ represents the conditional correlation among $d$ objects which can be obtained as $\rho_{ij} = - \frac{\theta_{ij}}{\sqrt{\theta_{ii}} \sqrt{\theta_{jj}}}$, where $\theta_{ij}$ is the $ij$-th element of the precision matrix $\Theta$.

We aim to construct a configuration of $d$ data points in a one--dimensional Euclidean space by using information about the distances between the $d$ nodes. Given the distance matrix $D$, we define a linear ordering $L$ of $d$ elements such that the distance $\widehat{D}$ between them is similar to $D$. We consider a metric MDS, which minimizes $\widehat{L}= \mbox{arg}\min_L \sum\limits_{i=1}^{d}\sum\limits_{j=1}^{d}(D_{ij} - \widehat{D}_{ij})^2$ across all linear orderings.
\paragraph{Outbred} An outbred population derived from mating two non-homozygous parents results in markers in the genome of progenies that can not easily be assigned to their parental homologues. Neighboring markers that vary only on different haploids will appear as independent, therefore requiring a different ordering algorithm [see Figure \ref{schemeCI}c]. In that case, to order markers we use the reverse Cuthill-McKee (RCM) algorithm \citep{cuthill1969reducing}. This algorithm is based on graph models. It reduces the bandwidth of the associated adjacency matrix, $A_{d\times d}$, for the sparse matrix $\widehat{\Theta}_\lambda^{(d)}$. The bandwidth of the matrix $A$ is defined by $\beta = \max_ {\theta_{ij} \ne 0} | i - j|$. The RCM algorithm produces a permutation matrix $P$ such that $P A P^T$ has a smaller bandwidth than does $A$. The bandwidth is decreased by moving the non-zero elements of the matrix $A$ closer to the main diagonal. The way to move the non-zero elements is determined by relabeling the nodes in graph $G(V_d, E_d)$ in consecutive order. Moreover, all of the nonzero elements are clustered near the main diagonal. 

\begin{figure}[t!] 
\centering{
\includegraphics[width=0.7\textwidth]{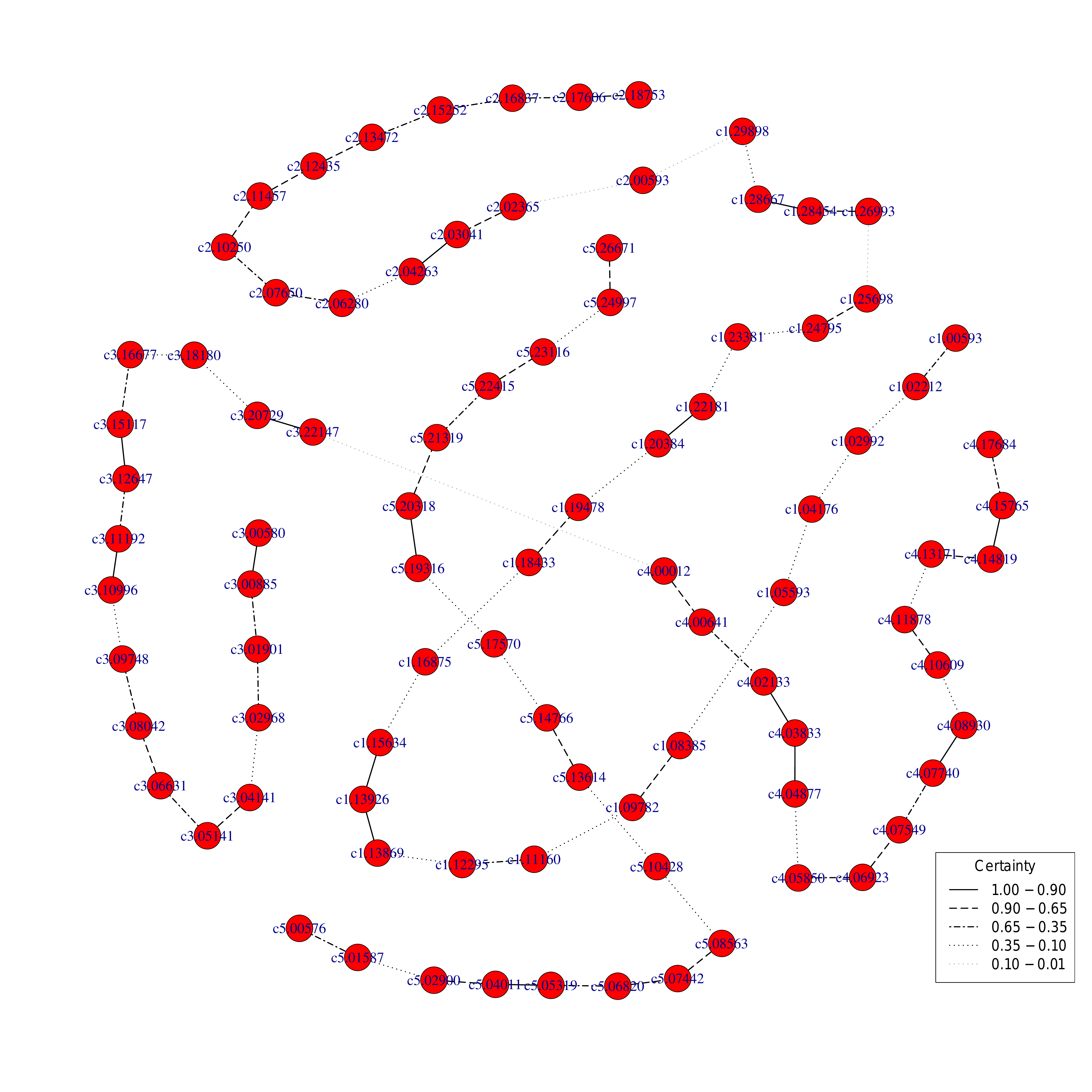} 
}
\caption{The certainty associated with the linkage map estimation in A.thaliana using the non-parametric bootstrap. }
\label{probExistingLinks}
\end{figure}

\subsection{Uncertainty in map construction}
\label{uncertainty}
Both empirical estimation of marginals and selection of the tuning parameter arise uncertainty in the map construction procedure. We compute the uncertainty associated with the estimated linkage map through a non-parametric bootstrap. We replicate $B$ datasets that are created by sampling with replacement $n$ samples from the dataset $Y_{n \times p}$. We run the entire map construction procedure to each bootstrap dataset. 
Each estimated map is associated with an adjacency matrix. The average of the B bootstrap adjacency matrices for the bootstrap samples reflects the underlying uncertainty in the estimation procedure of the linkage map construction.   

We have applied this procedure to evaluate the uncertainty associated with the estimation of the linkage map for the example data set $Cvi \times Col$ in \emph{A.thaliana}. This well-studied experiment is derived from a RIL cross between Columbia-0 (Col-0) and the Cape Verde Island (Cvi-0), where $367$ individual plants were genotyped across $90$ genetic markers \citep{simon2008qtl}. The $Cvi-0 \times Col-0$ RIL is a diploid population with three possible genotypes, where the genotypes are coded as $\{0, 1, 2\}$, where $0$ and $2$ represent two homozygous genotypes (AA resp. BB) from Col-0 and Cvi-0, and $1$ defines the heterozygous genotype (AB). 
We generate $100$ independent bootstrap samples from the Col-0 and Cvi-0 cross. For each $100$ bootstrap samples, we apply the map construction algorithm. Figure \ref{probExistingLinks} presents the certainty associated with the estimated linkage map for a subsample of $n= 50$ plants. The line type shows the estimated certainty associated with each link. 
For example, the gray dotted between marker ``c2.00593" from chromosome 2 and marker ``c1.298998" from chromosome 1 has the certainty value of 0.01, likewise for markers ``c3.22147" and ``c4.00012" from chromosomes 3 and 4. To sum up, for the links in the original dataset we obtain a $56\%$ certainty that the links are really there, whereas for the non-links in the original dataset we are $98\%$ certain that they are not there. 
We remark that when we use all $n = 367$ individuals, all the 100 bootstrap samples estimate an identical linkage map, which is the reason why for illustrative purpose, we used a subsample  $n = 50$.

\section{Simulation study}
In this section, we study the performance of the proposed method for different diploids and polyploids. In section \ref{simDiploid} we perform a comprehensive simulation study to compare the performance of the proposed algorithm with other available tools in diploid map constructions, namely J{\scriptsize OIN}M{\scriptsize AP} \citep{jansen2001constructing} and MSTMap \citep{wu2008efficient}. The former is based on Monte Carlo maximum likelihood and the latter uses a minimum spanning tree of a graph.

In section \ref{simPolyploid} we perform a simulation study to examine the algorithm performance on polyploids. At this moment the proposed method is the only one that constructs linkage maps for polyploid species automatically without any manual adjustment. Thus, in this case we can not compare the proposed method with other methods.
\subsection{Diploid species}
\label{simDiploid}
We simulate genotype data from an inbred $F2$ population. This population type generates discrete random variables with values $Y = \{0, 1, 2\}$ associated with the three distinct genotype states, $AA$, $Aa$, and $aa$ at each marker. The procedure in generating genotype data is as follows: first, two homozygous parental lines are simulated with genotypes AA and aa at each locus. A given number of markers, $p$, are spaced along the predefined chromosomes. Then, two parental lines are crossed to give an $F1$ population with all heterozygous genotypes $Aa$ at each marker location. Finally, a desired number of individuals, $n$, are simulated from the gametes produced by the $F1$ population. 

A genotyping error means that the observed genotype for an individual is not identical to its true genotype, for example, observing genotype AA when Aa is the true genotype. 
Genotyping errors can distort the final genetic map, especially by incorrectly ordering markers and inflating map length. Therefore, to order markers that contain genotyping errors is an essential task in constructing high-quality linkage maps. To investigate this, we create genotyping errors in the simulated datasets by randomly flipping the heterozygous loci along the chromosomes to either one of the homozygous allele. 

For each simulated data, we compare the performance of the map construction in {\tt netgwas} with two other models: J{\scriptsize OIN}M{\scriptsize AP}, and MSTMap. We compute two criteria: grouping accuracy (GA) and ordering accuracy (OA), to assess the performance of the above mentioned tools in estimating the correct map. The former measures the closeness of the estimated number of linkage groups to the correct number, and the latter calculates the ratio of markers that are correctly ordered. We define the grouping accuracy as follows: $ GA= \frac{1}{1+ (LG - \widehat{LG})^2 }$, where $LG$ stands for actual number of linkage groups and $\widehat{LG}$ is the estimated number of linkage groups. The GA criterion is a positive value with a maximum of $1$. A high value of GA indicates good performance in determining the correct number of linkage groups. To compute ordering accuracy, 
we calculate the Jaccard distance, $d_J$, which measures mismatches between the estimated order and the true order. We define the ordering accuracy of the estimated map as $
OA  = \frac{1} {1 + d_J}$. This measurement lies between $0$ and $1$, where $1$ and $0$ stand for a perfect and a poor ordering, respectively.

In terms of tion, {\tt netgwas} runs in parallel. In the performed simulations, we ran the map construction functions, both in {\tt netgwas} and the MST{\scriptsize MAP} on a Linux machine with 24 2.5 GHz Intel Xeon processors and 128 GB memory. J{\scriptsize OIN}M{\scriptsize AP} runs only on Windows. We ran it on a Windows machine with 3.20 GHz Intel Xeon processors and 8 GB RAM memory. 
\begin{figure}[t]
\centering
\includegraphics[width=1\textwidth]{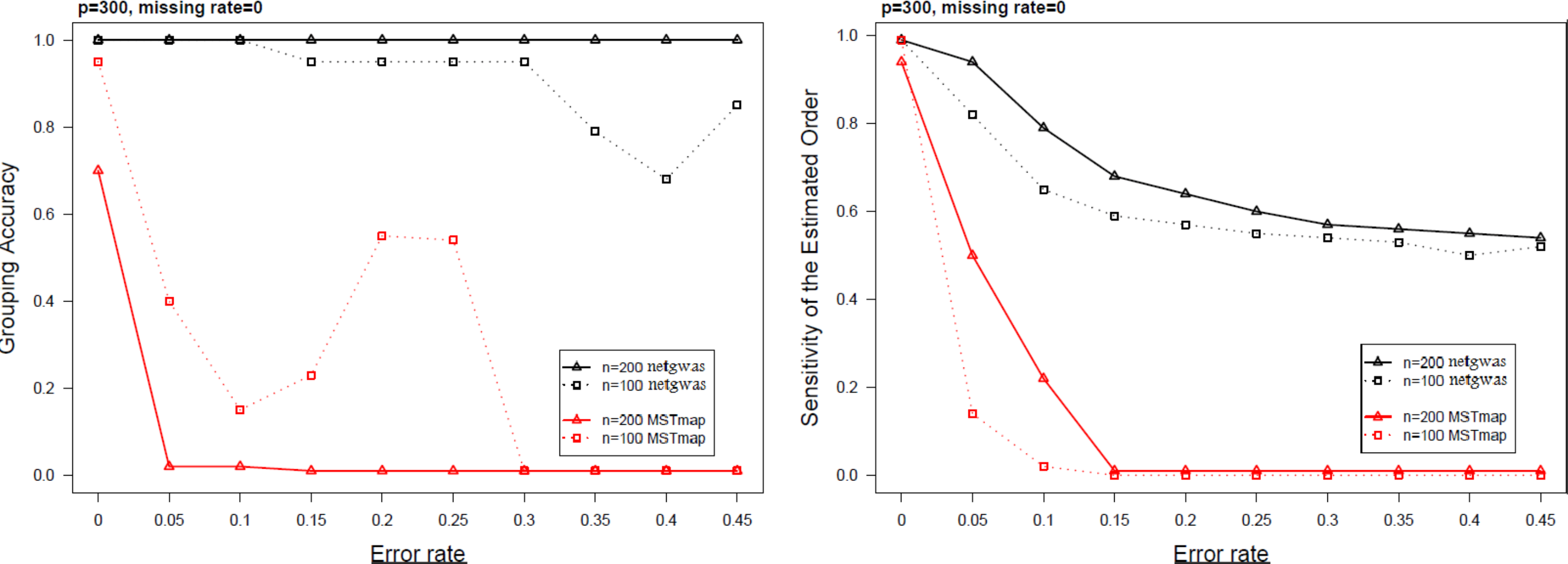}
(a) \hspace{7cm} (b)
\caption{Comparison of performance between map construction in {\tt netgwas} and MST{\scriptsize MAP} for different genotyping error rates. Variables $p$ and $n$ represent numbers of markers and individuals in simulated diploid genotype datasets. (a) Reports grouping, and (b) shows ordering accuracy scores for $50$ independent runs.}
\label{figure:ComparisonDifError}
\end{figure}
\paragraph{Evaluation of estimated maps in presence of genotyping errors} We studied the accuracy of the estimated linkage maps using two methods: {\tt netgwas} and MST{\scriptsize MAP} where genotyping errors are randomly distributed across the genetic markers. The simulated data contained $300$ markers for both $n=100$ and $n=200$ individuals where the genotyping error rates ranged from $0$ up to $0.45$. 
In these sets of simulations we activated the error-detection feature in MST{\scriptsize MAP}.

Figure \ref{figure:ComparisonDifError} evaluates the accuracy of estimated maps in terms of grouping (Figure \ref{figure:ComparisonDifError}a) and ordering accuracies (Figure \ref{figure:ComparisonDifError}b). In general, this figure shows that {\tt netgwas} constructed significantly better maps than MST{\scriptsize MAP} across the full range of genotyping error rates. More specifically, for a moderate number of individuals, $(n=200)$, Figure \ref{figure:ComparisonDifError}a shows that {\tt netgwas} correctly estimated the actual number of linkage groups for the all range of genotyping error rates. When $n=100$ {\tt netgwas} perfectly estimated the actual number of linkage groups up to $10\%$ genotyping errors, and very accurately ($\geq 0.95$) estimated the number of linkage groups for error rates between $10\%$ and $30\%$. With more than $30\%$ genotyping errors the accuracy diminished. MST{\scriptsize MAP} always made significantly poorer estimates of the actual number of linkage groups than did {\tt netgwas}; its performance immediately began to drop as soon as there was some level of genotyping errors. Surprisingly, it estimated the number of linkage groups better when $n=100$ than $n=200$, but this may have been a fluke.  

Figure \ref{figure:ComparisonDifError}b shows the ordering accuracy within each correctly estimated linkage group. Ordering quality in {\tt netgwas} was significantly better than MST{\scriptsize MAP} for both $n=100$ and $n=200$. This is because conditional independence is an effective way to recover relationships among genetic markers. More specifically, when $n=200$ and the error rate equaled zero, {\tt netgwas} ordered markers perfectly ($100\%$ accuracy) and MST{\scriptsize MAP} orders markers with a high accuracy $(95\%)$. In addition, with increased genotyping error rates, the map construction in {\tt netgwas} outperformed that of the MST{\scriptsize MAP} in ordering markers within each LG. Based on our simulations, we remark that with both {\tt netgwas} and MST{\scriptsize MAP} erroneous markers remain in the estimated linkage map. However, {\tt netgwas} orders them in the correct LG (see Figure \ref{figure:ComparisonDifError}), whereas MST{\scriptsize MAP} performs poorly in detecting LGs as well as in correctly ordering markers.
\begin{figure}[t]
\centering
\hspace*{-1.2cm}
\includegraphics[width=0.5\textwidth]{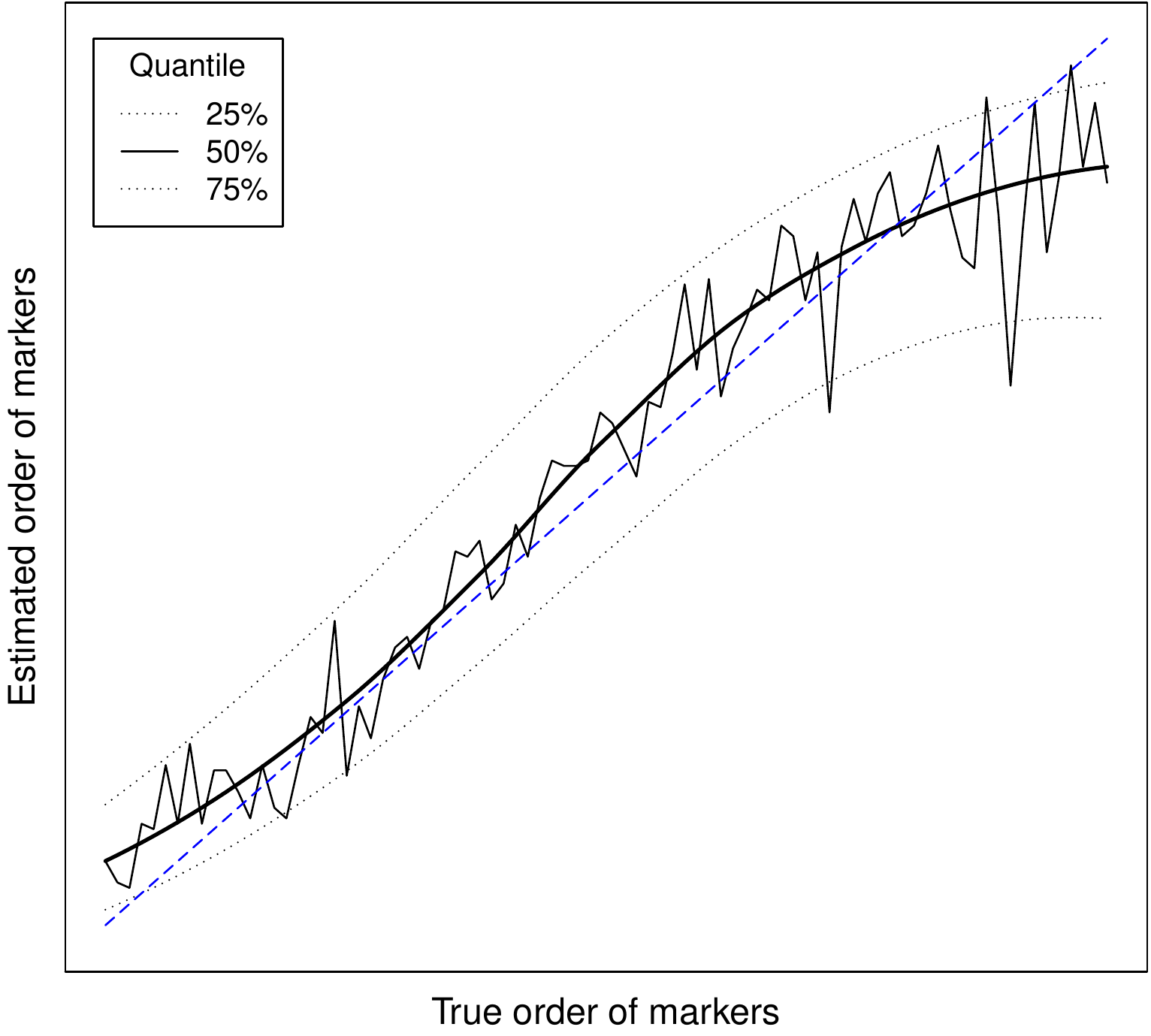} 
\includegraphics[width=0.5\textwidth]{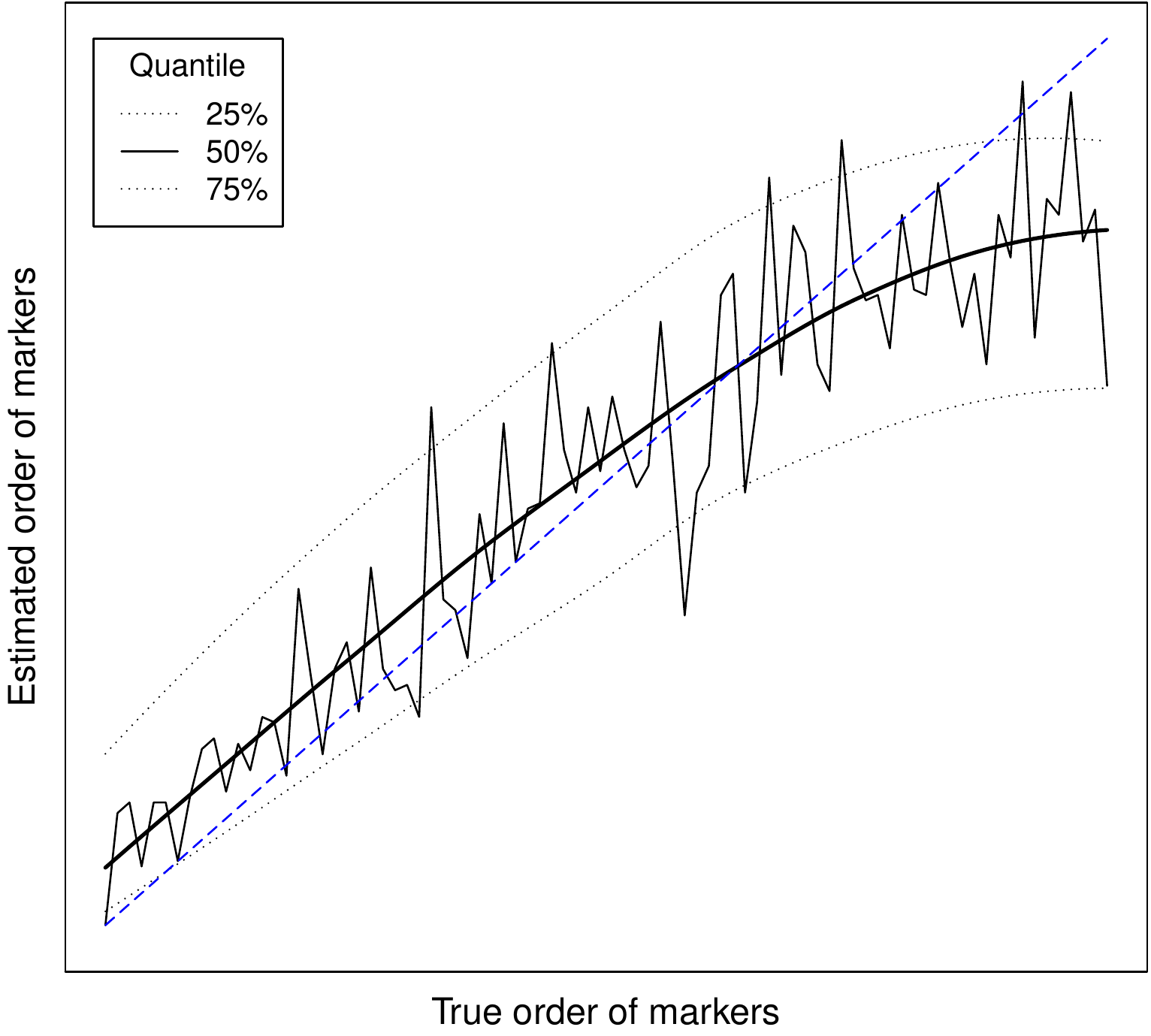}\\
 (a)\hspace*{8cm}(b)
\caption{Performance of {\tt netgwas} on different polypoid simulated datasets. Median, lower quartile, and upper quartile of estimated order versus true order for (a) tetraploids $(q = 4)$, and (b) hexaploid simulated datasets $(q = 6)$. Solid lines indicate median and smoothed median. Blue dashed line indicates ideal ordering.}
\label{ploidy}
\end{figure}
\subsection{Polyploid species}
\label{simPolyploid}
We also applied {\tt netgwas} to simulated outbred polyploid genotype datasets. We used PedigreeSim \citep{voorrips2012simulation} to simulate $F1$ mapping populations in tetraploids ($q= 4$) and hexaploids ($q= 6$) with $n = 200$ individuals. PedigreeSim simulates polyploid genotypes with different configurations, such as chromosomal pairing modes during meiosis. The simulated tetraploids ($q=4$) are motivated by autotetraploid potato where $Y = \{0, 1, 2, 3, 4\}$ corresponds to the five biallelic tetraploid genotype states $(aaaa$, $Aaaa$, $AAaa$, $AAAa$, $AAAA)$, which are created across $12$ chromosomes. The simulated hexaploids ($q=6$) are motivated by allohexaploid peanut, a polyploid species that contains $10$ chromosomes, where $Y= \{0, 1, 2, 3, 4, 5, 6 \}$ corresponds to the seven genotype states $(aaaaaa$, $Aaaaaa$, $AAaaaa$, $AAAaaa$, $AAAAaa$, $AAAAAa$, $AAAAAA)$ across its genome. In total, $50$ populations, each consisting of $p = 1000$ markers, were simulated for each scenario. 

We used the mean square error (MSE) as a measure for evaluating the performance of the proposed method on detecting the true number of chromosomes. In the tetraploid simulation the mean of MSE was $0.52$, and for the hexaploid simulation it was $0.15$. Figure \ref{ploidy} shows the performance of the proposed method in ordering markers for tetraploids (Figure \ref{ploidy}a) and hexaploids (Figure \ref{ploidy}b). The solid line shows the median of estimated order of each marker across a chromosome versus the true order, and the lower ($25\%$) and upper quartiles ($75\%$) of the estimated marker order is shown as dashed lines. This figure shows that, although ordering markers in outcrossing families is challenging [see section \ref{meioisMD}], the proposed method orders markers reasonably well. 
\begin{figure}[t]
\includegraphics[width=1\textwidth]{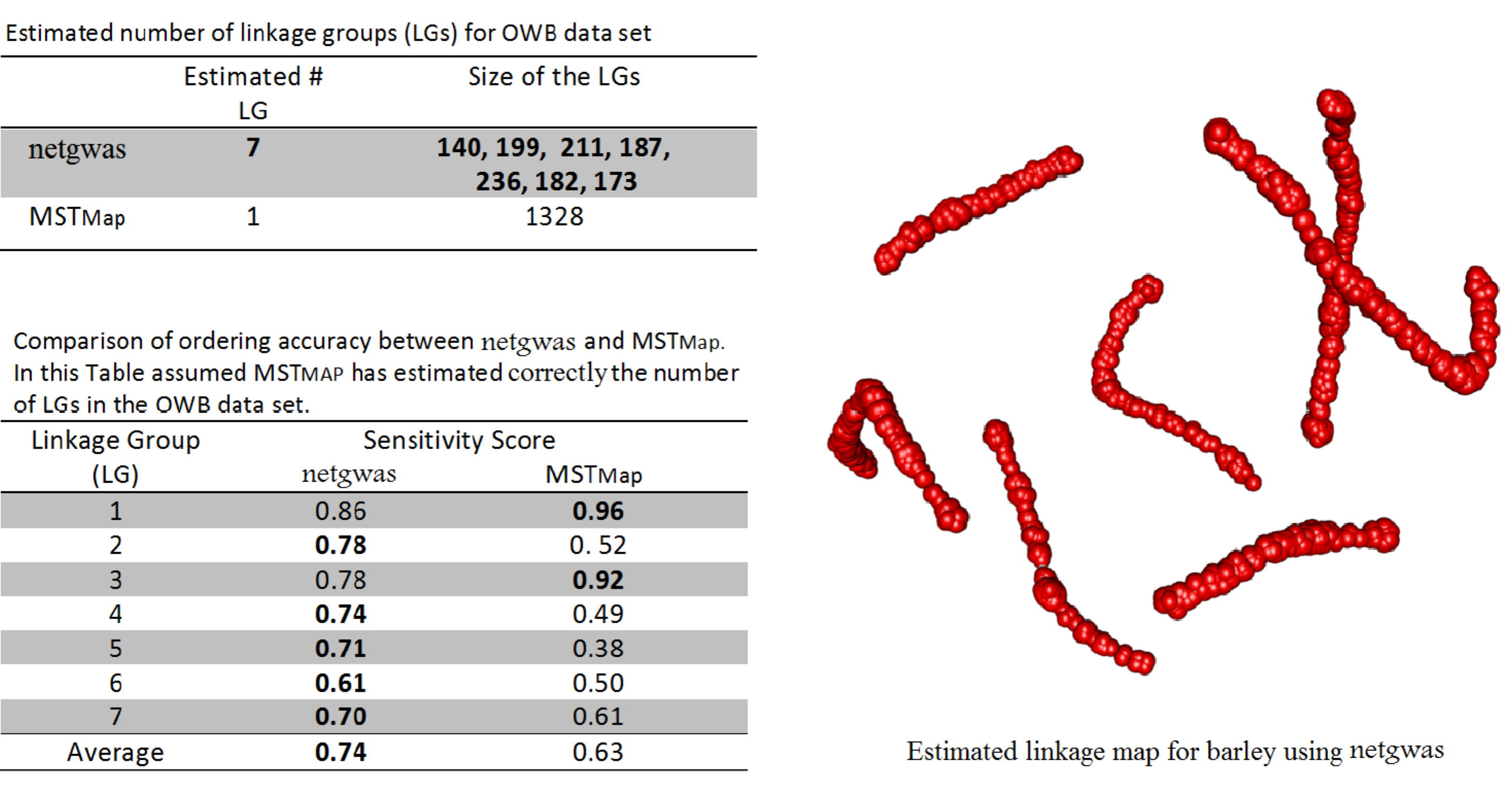}
\caption{Summary of comparison between {\tt netgwas} and MST{\scriptsize MAP} in barley data. Table summarizes estimated number of LGs (chromosomes) and size of markers within each LG. Below, average ordering accuracy scores for the two methods. Right figure estimated undirected graph in {\tt netgwas} for the barley data. This consists of $7$ sub--graphs, each showing a chromosome.}
\label{OWB}
\end{figure}

\section{Construction of linkage map for diploid barley}
In the literature a barley genotyping dataset is used to compare different map construction methods for real-world diploid data. This genotyping dataset is generated from a doubled haploid population, which results in homozygous individual plants, $Y_{ij} \in \{0,1\}$. 
Barley genotype data are the result of crossing Oregon Wolfe Barley Dominant with Oregon Wolfe Barley Recessive (see \url{http://wheat.pw.usda.gov/ggpages/maps/OWB}). The Oregon Wolfe Barley (OWB) data include $p = 1328$ markers that were genotyped on $n =175$ individuals of which $0.02\%$ genotypes are missing. The barley dataset is expected to yield $7$ linkage groups, one for each of the $7$ barley chromosomes. 

As shown in Figure \ref{OWB}, through estimating $\widehat{\Theta}_\lambda$, which contains conditional (in)\-de\-pen\-dence relationships between barley markers, we were able to correctly detect the $7$ barley chromosomes as sub--graphs in the estimated undirected graph. Furthermore, using the conditional correlation matrix as distance in the multi-dimensional scaling approach helped us to order markers with high accuracy. In addition, Figure \ref{OWB} reports the result of applying the two methods: {\tt netgwas} and MST{\scriptsize MAP}, to construct a linkage map for the barley data. The top part of Figure \ref{OWB} shows that our method correctly estimated the true number of chromosomes. Also, the size of markers within each chromosome is consistent with the number of markers that reported in \cite{cistue2011comparative}. MST{\scriptsize MAP} was not able to estimate the true number of chromosomes and grouped all $1328$ markers as one linkage group. The bottom of Figure \ref{OWB} shows the accuracy of estimated marker order in $7$ barley chromosomes. To be able to compare marker order in both methods we used the actual map to cluster markers in the map resulting from MST{\scriptsize MAP}. Thus, at the bottom of Figure \ref{OWB} it is assumed that the MST{\scriptsize MAP} has estimated the correct number of chromosomes. Average ordering of accuracy scores across the linkage groups in {\tt netgwas} is higher than those in MST{\scriptsize MAP} except with chromosomes 1 and 3.
\begin{figure}[t!]
\hspace{-0.8cm}
\includegraphics[width=0.46\textwidth]{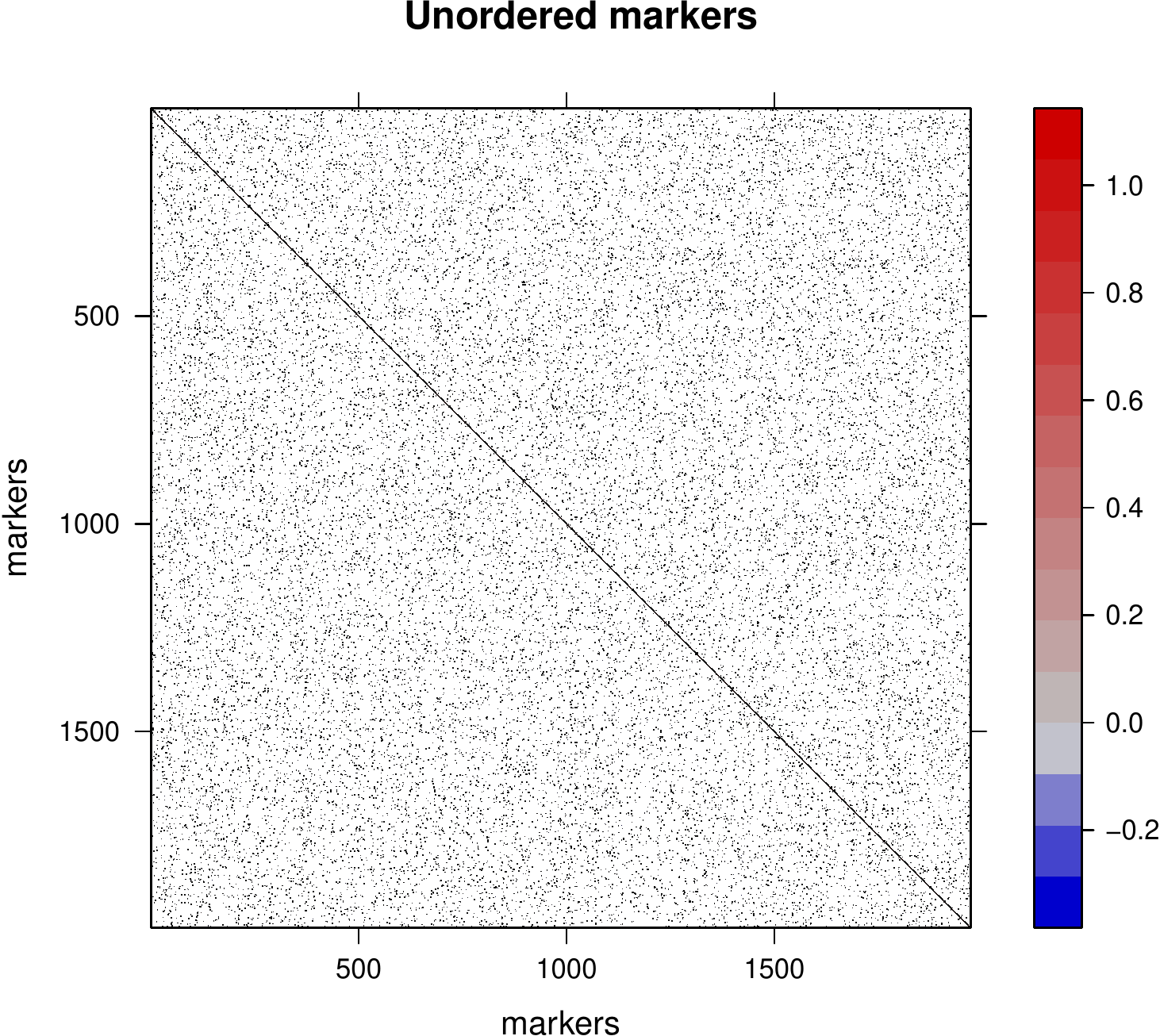} 
\hspace{-0.9cm}
\includegraphics[width=0.46\textwidth]{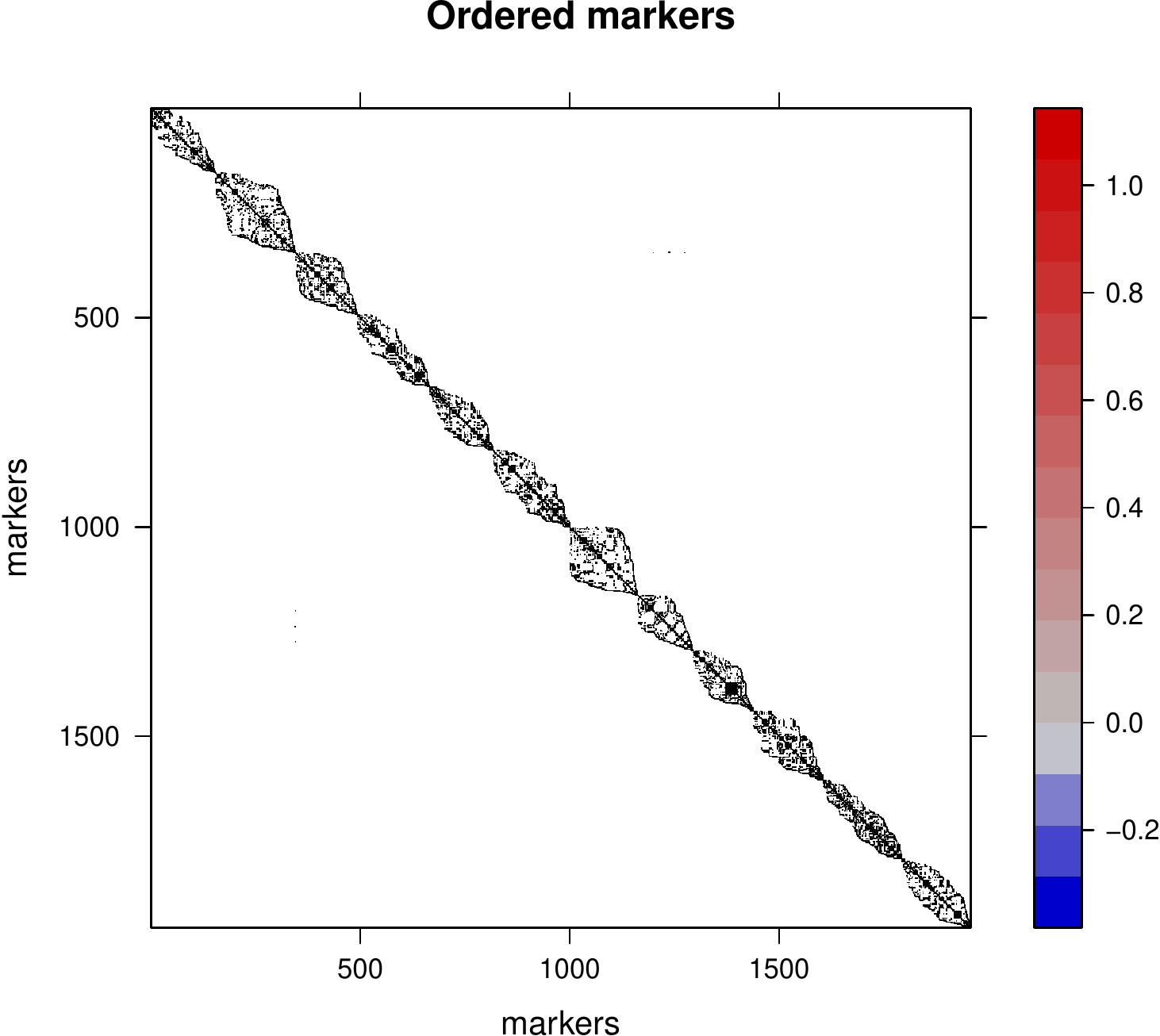} 
\hspace*{3.2cm} (a) \hspace{7.3cm}  (b)

\vspace{-0.8cm}
\centering
\includegraphics[width=0.48\textwidth]{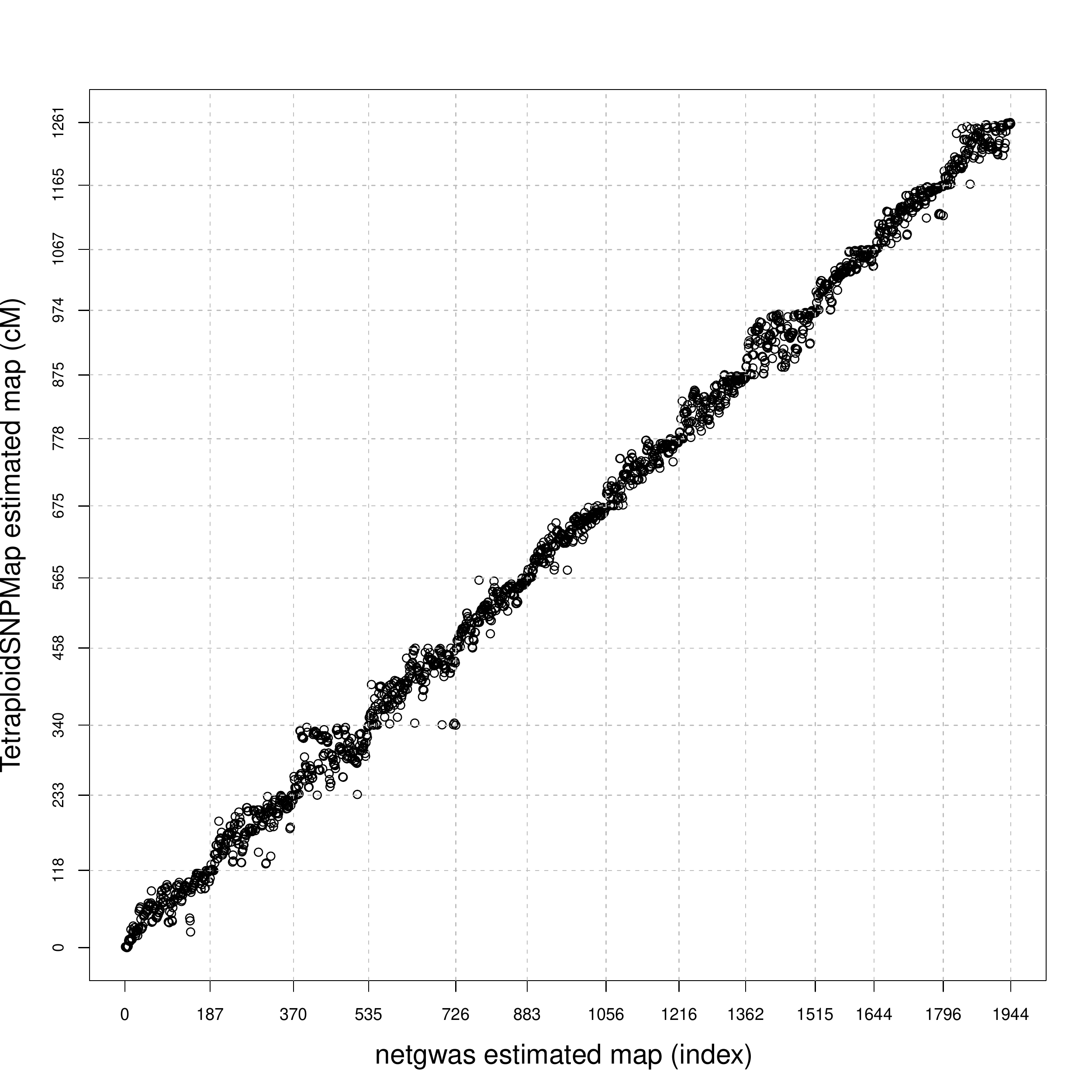} 

\hspace*{.5cm} (c)
\caption{Construction linkage map in potato. (a) Estimated precision matrix for 
unordered genotype data of tetraploid potato. (b) Estimated precision matrix after ordering markers. (c) Estimated order of markers across potato genome, versus estimated order in tetraploidSNPmap software. Each dashed line represents a chromosome. All potato chromosomes 
were detected correctly in netgwas.} 
\label{potato2015}
\end{figure}
\section{Construction of linkage map for tetraploid potato}
World-wide, the potato is the third most important food crop \citep{bradshaw2010potatoes}. However, the complex genetic structure of tetraploid potatoe's (Solanum tuberosum L.)  makes it difficult to improve important traits such as disease resistance in this crop. Thus there is a great interest in constructing linkage maps in the potato to identify markers related to disease resistance genes.

The full-sib mapping population MSL603 consists of $156$ F1 plants resulting from a cross between female parent “Jacqueline Lee” and male parent “MSG227-2”. The obtained genotype data contain $1972$ SNP markers \citep{massa2015genetic} with five allele dosages which are associated with the random variables $Y_j \in \{0,1, \ldots, 4\}$ for $j = 1, \ldots, 1971$. 

Figure \ref{potato2015} represents the result of applying the proposed map construction method to the unordered potato genotype data. Figure \ref{potato2015}a shows the estimated sparse precision matrix for the unordered genotype data. Figure \ref{potato2015}b represents the estimated precision matrix after ordering markers; it reveals the number of potato chromosomes as blocks across the diagonal. The potato genome contains $12$ chromosomes. The proposed method correctly identifies all $12$ chromosomes. The estimated linkage map contains $1957$ markers. Figure \ref{potato2015}c compares the estimated order in netgwas versus the estimated order in \cite{massa2015genetic} using the TetraploidSNPMap software \citep{hackett2017tetraploidsnpmap}. Each dashed line shows the estimated linkage group (LG), where netgwas estimates LGs using the eBIC criteria in (\ref{ebic}) and in TetraploidSNPMap the number of LGs should be specified manually. Given that the ordering of markers has always been a challenging task in linkage map constructions, and in particular for polyploid species, both methods ordered markers with 
similar precision except in chromosome 9 where TetraploidSNPMap suggests a different ordering. 

\section{Conclusion}
\label{conclusion}

Construction of linkage maps is a fundamental and necessary step for detailed genetic study of diseases and traits. A high-quality linkage map provides opportunities for greater throughput gene manipulation and phenotype improvement.
Here we have introduced a novel method for constructing linkage maps from high-throughput genotype data where the number of genetic markers exceeds the number of individuals. The proposed method makes full use of SNP dosage data to construct a linkage map for any bi--parental diploid or polyploid population. We propose to build linkage maps in two steps: (i) inferring conditional independence relationships between markers on the genome; (ii) ordering markers in each linkage group, typically a chromosome. In the first step of the proposed method we used the Markov properties of adjacent markers: the genotype of an individual haploid at marker $Y_j$ given its genotype at $Y_{j-1}$ or ${Y_{j+1}}$ is conditionally independent from the genotype at any other marker location. This property defines a graphical model for discrete random variables. 

We employed a Gaussian copula graphical model combined with a penalized EM algorithm to estimate a sparse precision matrix $\widehat{\Theta}_\lambda$. This method iteratively computes the conditional expectation of the complete penalized log-likelihood, and optimizes it to estimate $\widehat{\Theta}_\lambda$. The method can also deal with missing values, which are very common in genotype datasets. The nonparanormal skeptic is an alternative approach that is computationally faster but can not deal with missing genotypes. 
The number of linkage groups is determined via the information criteria, eBIC. Detection of linkage groups in the existing map construction software is usually done by manual tuning; this, however, influences the output map, whereas our method detects linkage groups automatically in a data--driven way. 

Depending on the type of mapping population, inbred or outbred, we use either a multi-dimensional scaling approach or the Cuthill-McKee algorithm, respectively, in step $2$ of the proposed linkage map construction. Both ordering algorithms result in a one-dimensional map. We noted that in outcrossing populations it is difficult to order markers because a clear definition of the parental genotype is lacking.

We performed several simulation studies to compare the performance of the proposed method with other commonly used diploid map construction tools. To address the challenges in the construction of a linkage map from genotype data, we studied the performance of the proposed method on simulated data with high ratios of genotyping errors. 
As shown in our simulation studies, our method, called {\tt netgwas}, outperformed the commonly available linkage map tools, when the input data were noisy. 

As outlined in \cite{cervantes2008development}, constructing linkage maps in polyploids, with outcrossing behavior, is a challenging task. So far, based on our experience, no method has been developed to construct polyploid linkage maps for a large number of different marker types without any manual adjustment and/or visual inspection. Based on the simulated polyploids with outcrossing behavior, the proposed method detected the true number of linkage groups with high accuracy, and ordered markers with reasonable precision. 

We applied the proposed method to two genotype studies involving barley and potato. In the barley map construction, we correctly detected its $7$ chromosomes, whereas  other method grouped all  markers in one linkage group. The {\tt netgwas} method ordered markers with higher accuracy in most of the chromosomes. The method detected all the potato chromosomes, although it identified chromosome $10$ as two linkage groups. Its ordering of markers within each chromosome was a substantial improvement of what has been possible up until now. We remark that the proposed map construction method uses all possible marker types, unlike the other map construction methods, which use a subset of markers \citep{grandke2017pergola}.


We point out that {\tt netgwas} also works for multi-allelic loci, which are locations in a genome that contain three or more observed alleles. For example, assume that A, T, and G are three possible alleles at location $j$ on a genome, unlike the most usual cases whereby only two alleles can be observed at a location (e.g. A and G). We propose to analyze either separately or jointly a dataset containing multi-allelic loci. In the former case, observed alleles count once as reference, and therefore allow for one separate dataset. In the above example three datasets will be generated: the first dataset counts the number of A alleles as a reference, the second dataset counts the number of T alleles as a reference, and the third dataset counts the G allele as a reference. Each dataset can be analyzed separately; to control similarity between the estimated precision matrices the fused graphical lasso can be used. The final map can be obtained through ordering markers in an estimated precision matrix. In the latter case, in the example above, we combine all three datasets as one dataset in such a way that it creates three replicates of $n \times p$ dimension. Moreover, we analyze the obtained dataset and construct the final linkage map.
While modern sequencing methods may be able to create accurate physical maps, there is an important role for methods such as ours that creates a genetic map.
A physical map displays all the nucleotides (ATCG) of a chromosome, which defines the physical distance (base-pair) between markers, but they do not give information on the recombination rates between markers. 
Two markers may be very close genetically, i.e., very little recombination occurs between them, but very far apart physically. 

\bibliographystyle{Chicago}
\bibliography{ref}

\newpage
\end{document}